# How to Work a Crowd: Developing Crowd Capital through Crowdsourcing

John Prpic, Prashant P. Shukla, Jan H. Kietzmann, Ian P. McCarthy

## Abstract

Traditionally, the term 'crowd' was used almost exclusively in the context of people who self-organized around a common purpose, emotion, or experience. Today, however, firms often refer to crowds in discussions of how collections of individuals can be engaged for organizational purposes. Crowdsourcing–defined here as the use of information technologies to outsource business responsibilities to crowds–can now significantly influence a firm's ability to leverage previously unattainable resources to build competitive advantage. Nonetheless, many managers are hesitant to consider crowdsourcing because they do not understand how its various types can add value to the firm. In response, we explain what crowdsourcing is, the advantages it offers, and how firms can pursue crowdsourcing. We begin by formulating a crowdsourcing typology and show how its four categories—crowd voting, micro-task, idea, and solution crowdsourcing—can help firms develop 'crowd capital,' an organizational-level resource harnessed from the crowd. We then present a three-step process model for generating crowd capital. Step one includes important considerations that shape how a crowd is to be constructed. Step two outlines the capabilities firms need to develop to acquire and assimilate resources (e.g., knowledge, labor, funds) from the crowd. Step three outlines key decision areas that executives need to address to effectively engage crowds.

## 1. Crowds and Crowdsourcing

Not too long ago, the term 'crowd' was used almost exclusively in the context of people who self-organized around a common purpose, emotion, or experience. Crowds were sometimes seen as a positive occurrence– for instance, when they formed for political rallies or to support sports teams— but were more often associated negatively with riots, a mob mentality, or looting. Under today's lens, they are viewed more positively (Wexler, 2011). Crowds have become useful! It all started in 2006, when crowdsourcing was introduced as ''taking a function once performed by employees and outsourcing it to an undefined (and generally large) network of people in the form of an "open call'' (Howe, 2006, p. 1). The underlying concept of crowdsourcing, a combination of crowd and outsourcing, is that many hands make light work and that wisdom can be gleaned from crowds (Surowiecki, 2005) to overcome groupthink, leading to superior results (Majchrzak & Malhotra, 2013). Of course, such ambitions are not new, and organizations have long desired to make the most of dispersed knowledge whereby each individual has certain knowledge advantages over every other (Hayek, 1945). Though examples of using crowds to harness what is desired are abundant (for an interesting application, see Table 1), until recently, accessing and harnessing such resources at scale has been nearly impossible for organizations. Due in large part to the proliferation of the Internet, mobile technologies, and the recent explosion of social media (Kietzmann, Hermkens, McCarthy, &

Silvestre, 2011), organizations today are in a much better position to engage distributed crowds (Lakhani & Panetta, 2007) of individuals for their innovation and problem-solving needs (Afuah & Tucci, 2012; Boudreau & Lakhani, 2013).

As a result, more and more executives——from small startups to Fortune 500 companies alike——are trying to figure out what crowdsourcing really is, the bene- fits it can offer, and the processes they should follow to engage a crowd. In this formative stage of crowdsourcing, multiple streams of academic and practitioner-based literature——each using their own language——are developing independently of one an- other, without a unifying framework to understand the burgeoning phenomenon of crowd engagement. For executives who would like to explore crowd-based opportunities, this presents a multitude of options and possibilities, but also difficulties. One problem entails lack of a clear understanding of crowds , the various forms they can take, and the value they can offer. Another problem entails absence of a well-defined process to engage crowds. As a result, many executives are unable to develop strategies or are hesitant to allocate resources to crowdsourcing, resulting in missed opportunities for new competitive advantages resulting from engaging crowds.

To help provide clarity, we submit an overview of the different types of crowdsourcing. Then we introduce the crowd capital framework, supplying a systematic template for executives to recognize the value of information from crowds, therein mapping the steps to acquire and assimilate resources from crowds. Finally, we discuss the unique benefits that can be gained from crowds before concluding with some advice on how to best 'work a crowd.'

## 2. Types of crowdsourcing

Crowdsourcing as an online, distributed problem- solving model (Brabham, 2008) suggests that approaching crowds and asking for contributions can help organizations develop solutions to a variety of business challenges. In this context, the crowd is often treated as a single construct: a general col- lection of people that can be targeted by firms. However, just as organizations and their problems vary, so do the types of crowds and the different kinds of contributions they can offer the firm. The following typology of crowdsourcing suggests that managers can begin by identifying a business problem and then working outward from there, considering (1) what type of contributions are required from members of the crowd and (2) how these contributions will collectively help find a solution to their business problem.

First, the types of contributions required from the crowd could either call for specific objective contributions or for subjective content. Specific objective contributions help to achieve an impartial and unbiased result; here, bare facts matter and crowds can help find or create them. Subjective content contributions revolve around the judgments, opinions, perceptions, and beliefs of individuals in a crowd that are sought to collectively help solve a problem that calls for a subjective result.

Second, contributions need to be processed collectively to add value. Depending on the problem to be solved, the contributions must either be aggregated or filtered. Under aggregation, contributions collectively yield value when they are simply combined at face value to inform a decision, without requiring any prior validation. For instance, political elections call for people to express their choices via electoral ballots, which are then tallied to calculate the sums and averages of their collective preferences; the reasons for their choices are not important at this stage. Other problems, however, are more complex and call for crowd contributions to be qualitatively evaluated and filtered before being considered on their relative merits (e.g., when politicians invite constituents' opinions be- fore campaigning). Together, these two dimensions help executives distinguish among and understand the variety of crowdsourcing alternatives that exist today (see Figure 1).

Two forms of crowdsourcing rely on aggregation as the primary process: crowd voting and micro-task crowdsourcing. In crowd voting, organizations pose an issue to a crowd and aggregate the subjective responses derived from crowd participants to make a decision. Consider the popular television show American Idol, which allows viewers to support their preferred contestants by submitting votes online or via telephone or text. These votes are tallied at the end of the show and contestants with the fewest votes are eliminated from the competition. Similarly, so-called prediction markets (Arrow et al., 2008) activate the wisdom of the crowd through crowd voting. But rather than simply adding up votes, these markets arrive at specific predictions that can exceed the accuracy of experts by averaging the independent responses of crowd participants. For instance, Starwood Hotels and Resorts utilized an internal prediction market by asking a crowd of its own employees to select the best choice among a variety of potential marketing campaigns (Barlow, 2008).

In micro-task crowdsourcing, organizations engage a crowd to undertake work that is often un- achievable through standard procedures due to its sheer size or complexity. An organization may need to assemble a large data set, have numerous photos labeled and tagged, translate documents, or transcribe audio transcripts. Breaking such work into micro-tasks (Gino & Staats, 2012) allows daunting undertakings to be completed more quickly, cheaply, and efficiently. Consider how Google uses re- CAPTCHA (von Ahn, Maurer, McMillen, Abraham, & Blum, 2008) and the little––—and admittedly annoying––—dialogue boxes that ask users to enter the text snippets they see of distorted images on- screen. It is commonly believed that this web utility is only for authenticating human users, thus keeping websites from spambots. However, every time the task of entering characters is completed, individuals are actually digitizing what optical character recognition (OCR) software has been unable to read. In this way, micro-task crowdsourcing is helping to digitize the archives of The New York Times and moving old manuscripts into Google Books. Similarly, crowdfunding (Stemler, 2013) endeavors are a form of micro-task crowdsourcing whereby an overall highly ambitious financial goal is broken into smaller funding tasks and contributions consist of objective resources (herein 'funds') that are simply aggregated for each venture.

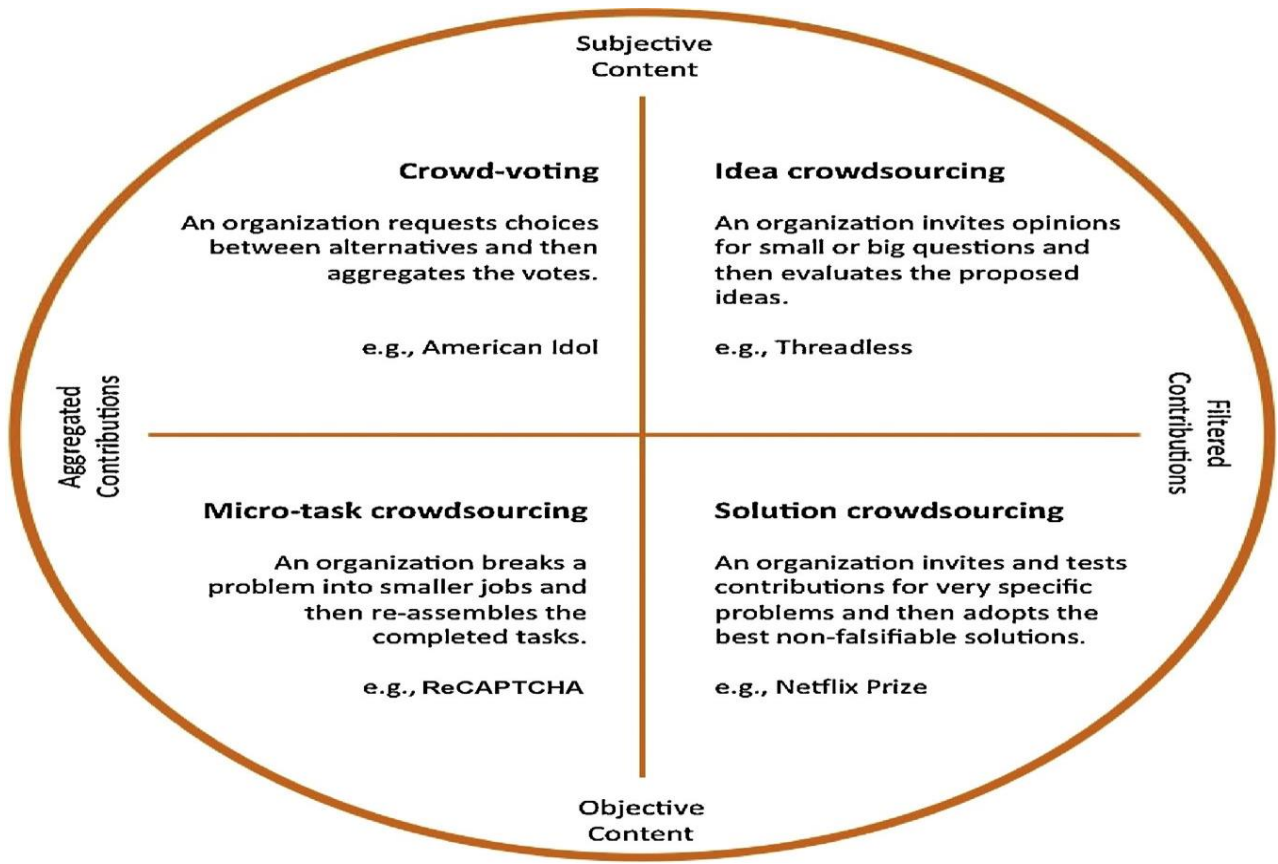

Figure 1. Crowdsourcing alternatives

Whether objective or subjective, crowdsourced contributions must be processed to be valuable. In idea crowdsourcing, organizations seek creativity from a crowd, hoping to leverage its diversity to generate unique solutions to problems/issues. An organization may receive many ideas from a crowd, which it will need to filter before one or more can be implemented. For instance, online artist community and e-commerce website Threadless asks the crowd for creative T-shirt designs and then internally chooses those it deems the most fitting ideas to be produced for sale (Brabham, 2010). Similarly, CineCoup seeks movie ideas in the form of trailers and then vets them, choosing which movie ideas will ultimately be financed for production (Fera, 2013).

In solution crowdsourcing–as opposed to idea crowdsourcing–organizations pose a well-defined and idiosyncratic problem to a crowd, potentially the organization's innovative and creative consumer base, asking for actual solutions (Berthon, Pitt, McCarthy, & Kates, 2007). Here, the organization can test, measure, and falsify solutions to deter- mine whether and to what degree the contribution actually solves the business problem. For instance, video streaming firm Netflix invited crowd members to participate in a competition to improve the company's predictive accuracy regarding how much viewers are going to enjoy a movie based on their extant film preferences (Bell & Koren, 2007; Zhou, Wilkinson, Schreiber, & Pan, 2008). Based on past data, the contributions were tested for accuracy, and the most effective solution won.

As the aforementioned forms of crowdsourcing produce a variety of potentials, these options can be implemented for differing goals. It is important to note that the different types of crowdsourcing can be implemented simultaneously or in a complementary fashion––—as organizational needs dictate––—as a crowdsourcing mix. Starwood Hotels and Resorts actually implemented an idea-crowdsourcing activity first, via which employees submitted different marketing campaign ideas, before the company used crowd voting to then select the best of the submitted marketing campaign ideas.

## 3. A crowd capital perspective

For any and all of the aforementioned initiatives, firms build crowd capital: organizational resources acquired through crowdsourcing. But this does not happen by accident; crowd capital is gained when the organization develops and follows a top-down process to seek bottom-up resources (e.g., knowledge, funds, opinions) from a crowd (Aitamurto, Leiponen, & Tee, 2011; Prpic & Shukla, 2013). In this section, we present this process as a three-stage model––—constructing a crowd, developing crowd capabilities, and harnessing crowd capital––—which offers unique benefits to executives seeking to enter the crowd milieu (see Figure 2).

Crowds need to be constructed––—they hardly ever pre-exist––—so in the first subsection that follows, we offer a detailed discussion of the important aspects to consider when constructing a crowd. Then, we describe crowd capabilities and summarize the two distinct stages of how organizations must (1) acquire content from a crowd and (2) assimilate the crowd-derived content into organizational practices (adapted from Zahra & George, 2002). Finally, we illustrate how constructed crowds and crowd capabilities can lead to the generation of crowd capital, and discuss the unique benefits this re- source can bring to organizations.

### 3.1. Constructing a crowd

Traditionally, executives and managers have worked with groups of individuals who are under direct control of the organization, for example in workgroups and project teams. These are relatively comfortable environments that do not involve dealing with strangers. More recently, organizations have also started to accept and appreciate contributions from groups that are outside of their direct control, consisting of people who span the boundaries of the organization–— for example, in communities that are virtual or mobile (Kietzmann et al., 2013). Regarding crowds, we are now asking executives to rethink who can add value to the organization, and how.

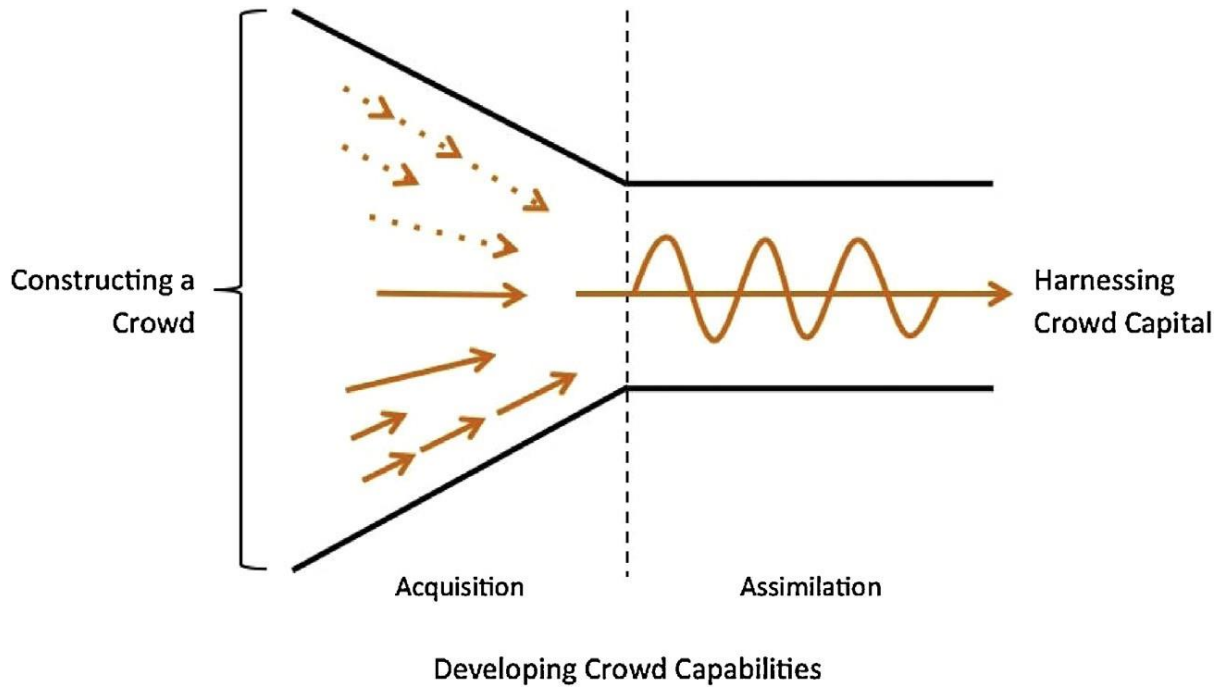

**Figure 2.** The crowd capital perspective

Although it might appear timely and considerate, and may send the right signals to organizational stakeholders, reaching out to crowds can only be of strategic value once the primary purpose for engaging a crowd is well aligned with organizational goals. Assuming that such an alignment is in place, firms next need to evaluate where the necessary contributions can be found; in other words, define the members of the crowd it wants to access (Frey, Luthje, & Haag, 2011) so the primary purpose of the activity can be achieved. Generally speaking, in terms of crowd size, larger scale——and thus large sample size, too——is thought to be beneficial, though scale alone is not the only consideration. Executives must also consider more narrowly where the solutions to their problems could come from. Should crowd members be derived solely from people out- side of the organization (e.g., to obtain new ideas) or from employees (i.e., to harness knowledge that already exists within the organization)? Further- more, should the crowd be accessible to anyone within these different populations or closed to selected types of participants? In some cases, no special talents are required and everyone's contributions can help perform organizational functions. In the micro-task crowd- sourcing example of Google's reCAPTCHA, anyone can complete this routine task of reading and entering characters from a screen. As a result, the crowd can decipher and enter about 30 million squiggles per day. In other situations, the crowd is more restricted and targeted at individuals who fulfill specific requirements or satisfy certain conditions. For example, Barclay's Bank assembled an external, closed group to help with development of the BarclayCard Ring (Marquit, 2013). Existing credit cardholders were invited to participate, narrow down, and vote on the terms and conditions associated with the new credit card.

Organizations can also construct crowds constituted of their own employees. Consumer electronics retailer Best Buy instituted a company-wide prediction market to forecast the success of new product ideas (Dvorak, 2008). But a crowd constituting an organization's own employees need not include the entire community. For example, when the U.S. Army launched ArmyCoCreate, a platform to canvass ideas for its Rapid Equipping Force, it actually did not invite all soldiers or officers, but rather only soldiers in the field at that time (Moore, 2014; Schiller, 2014). Clearly, these members of the crowd were very selectively invited from a closed, internal community. In this way, the U.S. Army tapped into only a section of its employee community for relevant knowledge and expertise, with the goal of solving real-life soldier challenges. Overall, the implications of these differing sources of crowds are clear. Different crowds possess different knowledge, skills, or other resources and, accordingly, can bring different types of value to an organization. Therefore, crowd construction is absolutely non-trivial in generating crowd capital.

### 3.2. Developing crowd capabilities

Implicit in the preceding discussion is the notion that an organization recognizes and is receptive to the value of resources dispersed in crowds. After an initial type of crowdsourcing is determined (the why) and the crowd construction is completed (the who), the organization needs to decide how it can (1) obtain resources dispersed in a crowd and (2) align crowd contributions with its existing internal processes. Working from the well-established absorptive capacity framework (Zahra & George, 2002), we refer to these two capabilities as acquisition and assimilation, respectively. Together, they comprise an organization's crowd capability (Prpic & Shukla, 2014).

### 3.2.1. Acquisition capabilities

Acquisition capabilities refer to an organization's proficiency in identifying and acquiring external resources that are useful toward its operations. In a crowd context, this capability mainly consists of (1) understanding the type of interaction that is required for the acquisition of knowledge and (2) choosing an appropriate IT structure that will facilitate the engagement of dispersed individuals in a crowd.

Different types of problems require different types of interaction between the crowd and the organization, and among individuals in the crowd itself. Regarding the former, the choices are related to those presented in the customer service literature, wherein conceptually distinct social mechanisms are used in the interaction between a customer and a firm (Gutek, Bhappu, Liao-Troth, & Cherry, 1999; Gutek, Groth, & Cherry, 2002). Any organization engaging a crowd needs to determine if crowd engagement should be based on encounters—— that is, on discrete transactions that could be repeated but are essentially independent——or on relationships, via which the organization and crowd members expect to have continued contact with one another in the future, possibly with no end in sight.

The second dimension of crowd interaction for knowledge acquisition relates to whether the individuals in a crowd need to interact with one another to generate the desired output. Should they work together on solving a problem through collaboration, or should the engagement be autonomous, via which individuals are not affiliated with one another and complete tasks independently?

These choices matter a great deal, as these two dimensions of interaction together influence the incentives for motivating individuals to participate and the choices of an appropriate IT structure. For instance, consider reCAPTCHA again as an example. Because quick episodes of interaction happen independently from Google and independently of other crowd members, human contributors volunteer about 83,000 hours of daily labor to Google, scanning documents while at the same time safe-guarding websites from spambots. The key insight from this acquisition capability is the realization that participants do not need to invest any time in understanding how to work with the organization or with each other. At the other extreme, enterprise wiki technologies (Jackson & Klobas, 2013) and enterprise social media (Mathiesen & Fielt, 2013) are based on relationships: ongoing cooperation is required to create and negotiate rich content from dispersed knowledge. The power of such a capability can also be considerable; for example, consider Best Buy, whose implementation of its solution crowd-sourcing tool Blue Shirt Nation wiki (Dvorak, 2008) connects 24,000 employees, allowing them to individually raise and discuss issues important to internal operations and to share customer service tips. Use of this form of interaction has allowed the company to quickly reverse internal policies that reduce employee morale. In sum, the type of crowd interaction chosen is a distinct design choice avail- able to the organization and has significant ramifications for organization-crowd dynamics. Given that crowdsourcing is almost always an IT- mediated activity, the choice of technology flows very much from earlier strategic decisions. The combination of the primary purpose of the activity (crowd voting, idea, micro-task, or solution crowd- sourcing), the boundaries of the crowd (inside or outside the organization or a mixture), and the type of interaction of participants with the organization (encounter or relationship, collaborative or autonomous) all heavily influence the chosen IT structure.

The vast majority of crowd-engaging IT employs a web-based or mobile platform, or uses the two in concert. These IT choices often start with the fundamental question of whether an organization should make or buy the technology it requires. There are always advantages and disadvantages for either choice pertaining to things such as quality and feature control, security, development cost, risk and time to market, IP ownership, and product maintenance. In this respect, crowdsourcing strategies are no exception, and organizations can choose between developing their own proprietary solutions or opting to operate through intermediaries.

In the realm of intermediaries, many offerings already exist, organized solely to help an organization generate crowd capital. One class of crowd capital facilitating intermediaries includes web- based spot-labor pools for micro-task crowdsourcing, such as Amazon's M-Turk (Buhrmester, Kwang, & Gosling, 2011; Little, Chilton, Goldman, & Miller, 2009), CrowdFlower (Biewald, 2012; Finin et al., 2010), and Samasource (Biewald, 2012; Nesbit & Janah, 2010).

These intermediaries have already cultivated large populations of participants, allowing organizations to quickly tap into a ready, willing, and able supply of affordable labor. In the case of Samasource and some other social enterprises (Seelos & Mair, 2005), the labor pool is sourced from the developing world, so using such an alternative to generate crowd capital may also serve an organization's corporate social responsibility goals via what Gino and Staats (2012) term 'impact sourcing.'

Another class of crowd capital facilitating intermediaries includes web-based 'tournament- style' intermediaries for solution and idea crowd- sourcing (Afuah & Tucci, 2012; Boudreau & Lakhani,2013), such as Innocentive, Eyeka, and Kaggle (Ben Taieb & Hyndman, 2014; Narayanan, Shi, & Rubinstein, 2011). Similar to the spot-labor pool sites such as M-Turk, sites like Innocentive and Kaggle have established a large pool of self-selected participants, though in these cases, the participants are problem solvers rather than workers for hire. Through these intermediaries an organization can post specific problems that need to be solved, and the intermediaries offer a variety of different packages and price points for the organization's problem-solving needs. These intermediaries can represent a highly successful strategy; consider Innocentive, for example, which boasts a cadre of 250,000 registered solvers and a success rate of greater than 50% (Aron, 2010).

### 3.2.2. Assimilation capabilities

As we have thus far outlined, an organization has many different decisions to consider before engaging a crowd through IT. However, we must emphasize that implementing all of the previous decisions success- fully does not guarantee generation of the desired crowd capital resource. Successfully engaging a crowd and acquiring the desired contributions from it are necessary, but not sufficient alone to generate crowd capital. The final element in the crowd capital creation process lies in the internal assimilation of crowd contributions. A separation between the acquisition and assimilation of crowd capital reflects arguments from organizational strategy scholars, who propose that value creation and value capture are two distinct processes (Lepak, Smith, & Taylor, 2007). Since the former does not naturally lead to the latter (e.g., Yahoo; Shafer, Smith, & Linder, 2005), value creation and value capture need to be considered individually and explicitly (Amit & Zott, 2001; Shafer et al., 2005). We proceed with a similar analogy and reason that both acquisition and assimilation strategies are independently important in the process of creating crowd capital.

As we have illustrated in Figure 1, some forms of crowd engagement require filtering and others re- quire aggregation of crowd contributions. In either case, organizations need to institute internal processes to organize and purpose the incoming knowledge and information. Such processes may include assigning the aggregation and filtering to specific teams within the organization or creating a new group concerned with the task. Depending on the goals of the endeavor, certain teams or individuals may be tasked with engaging individuals in the crowd to curate and manage the community, shaping crowd engagement and ensuring that desired contributions are elicited from the participants. Similarly, the organization should define a set of metrics before beginning crowd engagement to determine how success or failure might be evaluated. Such metrics can include measures for the size of the crowd, for contributions from the crowd, and/or for other tailored metrics specific to the endeavor. Research has shown that it may be advisable to assign experts in the specific field to interact with the crowd (Chun & Cho, 2012), and that it may be useful to determine ahead of time how the crowd contributions will be used within the organization (Brabham, 2012).

### 3.3. Harnessing crowd capital

As we pointed out in Section 2, it is useful to think of the different forms of crowdsourcing available to organizations as a mix whereby different types of crowdsourcing may be employed simultaneously or sequentially. Further, as noted in Section 4, knowledge is dispersed within the population of an organization. Bringing these two insights together, we suggest that–— depending upon the resource needs–— an organization can construct separate crowds as acquisition and assimilation capabilities. For example, an organization could construct a crowd comprised of its own employees as the filtering or aggregation mechanism to process the knowledge acquired from an external crowd. In turn, it may also be that the reverse situation is also beneficial; we see no reason why crowds within organizations could not be used to derive a list of problems, which could then be posted to a crowd outside the organization at a place like Innocentive (Aron, 2010) to capture a diverse range of ideas or solutions. Therefore, in pursuit of crowd capital, executives should not think of these applications as siloed potentialities, but rather as hypothetically overlapping tools brought to bear in an overall crowdsourcing mix.

Irrespective of what a competitor might do to mimic another's crowd capability, the crowd capital resource that is gained through the goal-focused, thought-out process that we detail here is hard to replicate. This, of course, is particularly true for subjective contributions that are filtered by the organization, gleaning unique and idiosyncratic resources for the organization that can lead to competitive advantages, a potentially positive addition to any business model (Barney, 1991; Shafer et al., 2005).

In addition, crowd capital can be generated with- out collaboration, lowering investment in gaining this resource. As illustrated herein, crowd capital can be generated through encounters or relation- ships with the firm. Many examples–—such as Google's reCAPTCHA or Microsoft's Asirra (Aggarwal, 2012; von Ahn et al., 2008), the Iowa Electronic Prediction market (Arrow et al., 2008), and Foldit (Cooper et al., 2010)–—illustrate the power of using encounters to generate crowd capital. Therefore, generating crowd capital by engaging the dispersed knowledge of a crowd does not require a community of individuals or their continuous participation. When deciding whether to make, buy, or rent a crowd capability, organizations must consider if they need to construct an encounter or relation- ship-based crowd, or some combination thereof.

## 4. Final thoughts on how to work a crowd

This article offers contributions to both the research and practitioner communities. We hope that our typology -separating crowdsourcing by the subjective or objective content obtained from the crowd, and then either aggregated or filtered by the organization– will help scholars develop lenses appropriate for research on crowd voting, micro-task crowdsourcing, idea crowdsourcing, and solution crowdsourcing, respectively. Herein, we present the crowd capital perspective (which illustrates in testable form a generalized process model of crowd construction) as well as acquisition and assimilation capabilities, leading ultimately to different forms of crowd capital. It is our hope that this early work on a crowd- sourcing process will motivate other researchers to tease apart the different kinds of capabilities needed for different types of crowdsourcing, and to study in more detail the different types of crowd capital these can create. Furthermore, our work on crowdsourcing may have the potential to inform literature in other management areas. In particular, a firm's need to construct a crowd based on the similarity of its members is comparable to marketers' need to segment their markets: to divide a heterogeneous market into homogeneous groups (Wedel & Kamakura, 1999).

Future research on how firms form their crowds from an amorphous group of people outside their boundaries might inform segmentation practices (e.g., Yankelovich & Meer, 2006), and vice versa. For the practitioner community, we contribute by illustrating key decision areas that executives need to consider and address to effectively engage crowds through IT. For instance, for decision makers, the crowd typology provides a suitable starting point for understanding what problems can be crowdsourced and the types of responses crowd-sourcing will yield. Crowdsourcing capabilities, both in terms of acquisition and assimilation, provide dimensions and examples of IT structures and engagement options that we hope will prove practical for decision makers and their strategic development of crowdsourcing initiatives.

In review, we close with the fundamental considerations for generating and benefiting from crowd capital. The first topic an organization needs to investigate is the content to be acquired using its crowd capability. In other words, what problem or opportunity can/should be addressed by leveraging crowd knowledge? Does the problem call for a subjective or objective solution, and should crowd contributions be aggregated or filtered to yield optimal value for the firm? From here, the organization can begin to think about constructing the pertinent crowd (i.e., where crowd members should come from: internal, external, or both) and what form(s) of IT will be used to engage members of the crowd (i.e., in encounters or relationships). Should crowd members collaborate with each other or work as autonomous agents? Should the appropriate IT structure be made (in house), bought, or rented (through intermediaries) so that dispersed crowd- based resources can be accessed?

Overall, the powerful insights of Hayek (1945) from about 70 years ago could not be more pertinent and significant in this day and age. As new technologies allow firms to reach more and more individuals and crowds, access to dispersed knowledge will continue to improve, allowing managers not only to consider crowdsourcing for the solution of a variety of everyday problems, but also to build crowdsourcing into their organizational strategies and underlying business models. For vanguard businesses, this change has already arrived. We hope this article convinces others that working a crowd and developing crowd capital through crowdsourcing can play a significant role in creating and sustaining competitive advantage.